\documentclass[11pt,twoside]{article}  
\usepackage{apn3conf}

\begin{document}   

%
%
%
%

\title{Orientation effects in bipolar nebulae: Can disks do it?}

%

\author{Hektor Monteiro, Hugo E. Schwarz, and Ryan T. Peterson\altaffilmark{1}}
\affil{Cerro Tololo Inter-American Observatory, NOAO-AURA, Casilla 603, 
La Serena, Chile}

\altaffiltext{1}{REU program 2003 - CTIO-NOAO-AURA, Chile}


\contact{H. Monteiro}
\email{hmonteiro@ctio.noao.edu}

%
%
%
%
%

\paindex{Monteiro, H.}
\aindex{Schwarz, H.}     
\aindex{Peterson, R.}     

%
%

\authormark{Monteiro, Schwarz \& Peterson}

%

\keywords{planetary nebulae, symbiotic, dust, modeling, SED}


\begin{abstract}          

In this work we investigate how a circumstellar disk affects the
radiation emitted by an embedded star. We show correlations obtained
from broad-band observations of bipolar nebulae indicating that an
orientation effect is at play in these systems. The FIR radiation
relative to total radiation increases with inclination while the
NIR and BVR fractions decrease. This is an expected effect if we
consider the system as being made up of a dense dusty disk being
irradiated by a hot star. We calculate 2-D models to try and reproduce
the observed behavior with different disk and star configurations.

\end{abstract}


\section{Observations}

We collected a sample of bipolar planetary nebula and symbiotic nebula
from the literature that contained sufficient data to produce a
reasonabble spectral energy distribution (SED). After constructing the
SEDs we compared the luminosity emitted in each band relative to the
total luminosity as a function of inclination angle. The inclination
angles were determined by the three authors independently, being in
agreement within 10-15$^o$ for nearly all objects, with an exceptional
30$^o$ difference in a couple of cases.  In Figure\,1 we show the plot
with the relative luminosities for each band as a function of the
inclination. One can clearly see the increase in the relative FIR flux
as well as the decrease of JHK and BVR with inceasing inclination.

\begin{figure}
\begin{center}
\epsscale{.80}
\plotone{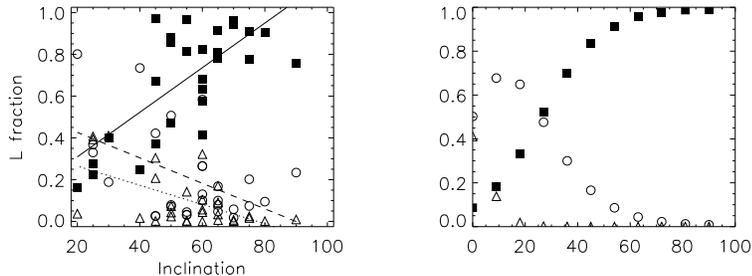}
\end{center}
\caption{Observed and model (for one system only) correlation between 
fractional luminosities and inclination angle.}
\end{figure}

\section{Model}

In an attempt to explain this observed behavior, we propose that the
systems are composed of a dusty disk being irradiated by a star or
binary. This radiation heats up the dust that is responsible for the
radiation emitted in the FIR. The combined effect of this re-radiation
with the extinction produced by the disk is what would create the
observed effect.

We use a simple 2-D radiative transfer model that calculates the
temperature of the grains of a disk defined within a square
grid. By requiring conservation of luminosity and iterating we
determine the temperature of the dust in each cell. Secondary
radiation is not yet accounted for.

For the dust characteristics we adopt typical interstellar grains with
R$_v$\,=\,3.1 with properties as calculated by Li \& Draine (2001) and
a gas to dust ratio of 125. For the disk we obtained temperature maps
from which we calculated a random sample of star-disk systems with
sin(i) distributed random inclination angles. This simple method
assumes a general temperature, taken here as the average temperature
obtained from the 2-D code. We also obtain attenuation functions from
the 2-D code that will be used to calculate the extinction of the
starlight when intercepted by the disk. Also from the code we obtain
the amount of starlight absorbed by the disk. For this component, we
assume that the energy is re-radiated in the FIR as a blackbody with
the average temperature mentioned before. Figure~1 shows one example
of such a model run.

\section{Conclusions}

The results show that a disk with sufficient density and size is able to
produce the general, observed behavior. It is interesting to note that the
simple model did not do well when we considered a double system with a
luminous cold star as well as the hot ionizing one. On the other hand, the
more precise 2-D code copes well with that binary and in fact needs it to
reproduce the observed results. These results are only preliminary and much
more detail should be accounted for to be able to pin-point the actual disk
structure needed. Even so, we can safely say that, yes, disks can do it!

%
%
%


\end{document}